From Plasticity to a Renormalisation Group

RC Ball*
R Blumenfeld*+

*Department of Physics, University of Warwick,
+Cavendish Laboratory, University of Cambridge

Abstract

The Marginally Rigid State is a candidate paradigm for what makes granular material a state of matter distinct from both liquid and solid. Coordination number is identified as a discriminating characteristic, and for rough-surfaced particles we show that the low values predicted are indeed approached in simple two dimensional experiments. We show calculations of the stress transmission suggesting that this is governed by local linear equations of constraint between the stress components. These constraints can in turn be related to the generalised forces conjugate to the motion of grains rolling over each other. The lack of a spatially coherent way of imposing a sign convention on these motions is a problem for up-scaling the equations, which leads us to attempt a renormalisation group calculation. Finally we discuss how perturbations propagate through such systems, suggesting a distinction between the behaviour of rough and of smooth grains.

Introduction

The transmission of stress through granular materials is an important practical issue, for example governing the safe design of foundations, embankments and silos. Some of its aspects can be quite counterintuitive, such as the liability of grain silos to collapse at the moment when grain is allowed to flow out from the base (just the opposite of expectation for a liquid), and measurements [1] showing that conical piles of granular material may distribute their gravitational load over the supporting base with a local minimum under the apex of the pile (just the opposite of expectation for a solid). Both these puzzles can be interpreted qualitatively [2] and the latter quantitatively [3] in terms of internal arching within a static pile transferring load toward the sides.

Marginally Rigid State

We propose here that as granular material first organises itself to come to rest, it should naturally attain a quite characteristic state which we term *Marginally Rigid* [4,5]. In doing so we will start by assuming the grains are ideally rigid objects [6], and that any given contact between grains is either perfectly smooth (permitting sliding and exchanging no tangential force) or perfectly rough with an infinite coefficient of static Coulomb friction. All of these assumptions we will revisit in due course. We will throughout assume that thermal fluctuations can be ignored.

The key idea is a simple counting argument. The system should consolidate until the number of degrees of freedom of the free grains is matched by the number of constraints dues to contacts between them. Now for every degree of freedom of the free grains there is precisely one corresponding balance condition of force (or torque) in order for the system to be in mechanical equilibrium, and for every constraint there is one intergranular force component to be determined. Thus consolidation should bring the system to the point where the conditions of force (and torque) balance are precisely sufficient to determine all the intergranular forces. This means the transmission of stress through the system is fully determined by the geometry of the contact network alone, without considering the internal behaviour of our grains – a surprising result which we suggest is the defining characteristic of the marginally rigid state.

The scope of this proposal becomes more clear when one considers the consequences in terms of mean coordination number for the resulting piles. In two dimensions, we have 3 degrees of freedom per grain and, if the grains are rough, then for each contact we have two constraints and as each contact contributes to the coordination of two grains the critical coordination number at which a stable assembly is first achieved is $z_c = 3$. This is a notably low value, which prompts experimental scrutiny below. By contrast when the grains are treated as ideally smooth (but not circular) there is only one constraint per contact, leading to $z_c = 6$ which is topologically the maximum possible in two dimensions, enabling us to assert that smooth grain assemblies can at most be marginally rigid. The results for three dimensions are analogous: rough grains give a low value, $z_c = 6$, whereas smooth aspherical grains give $z_c = 12$ which is hard (but not topologically impossible) to exceed. In both cases smooth particles of spherical symmetry are anomalous, as no torques can arise, leading to $z_c = 4$ and $z_c = 6$ in 2 and 3 dimensions respectively.

It is noteworthy that all of the smooth grain values for $z_c$ above match the corresponding values of coordination number achievable by sequential packing, if for aspherical grains one admits search of position and orientation to optimize coordination number.

Experimental Test

We identified coordination number 3 for rough grains in two dimensions as a 'smoking gun' test of the marginally rigid state, as we are not aware of any other perspective from which this result seems likely. We performed a somewhat idealized experiment in which grains initially stationary on a horizontal supporting surface were swept into a pile by a collector pushed slowly across the surface, as shown in figure 1. The stationary grains are in an analogue of (force)-free fall, whilst those swept into the advancing pile feel the analogue of a gravitational force due to sliding friction against the supporting surface. We worked slowly enough that inertia and momentum played negligible role, and simple calibration experiments indicate that the coefficient of static friction between our grains (which were cut from fibrous card) was at least as high as 50.

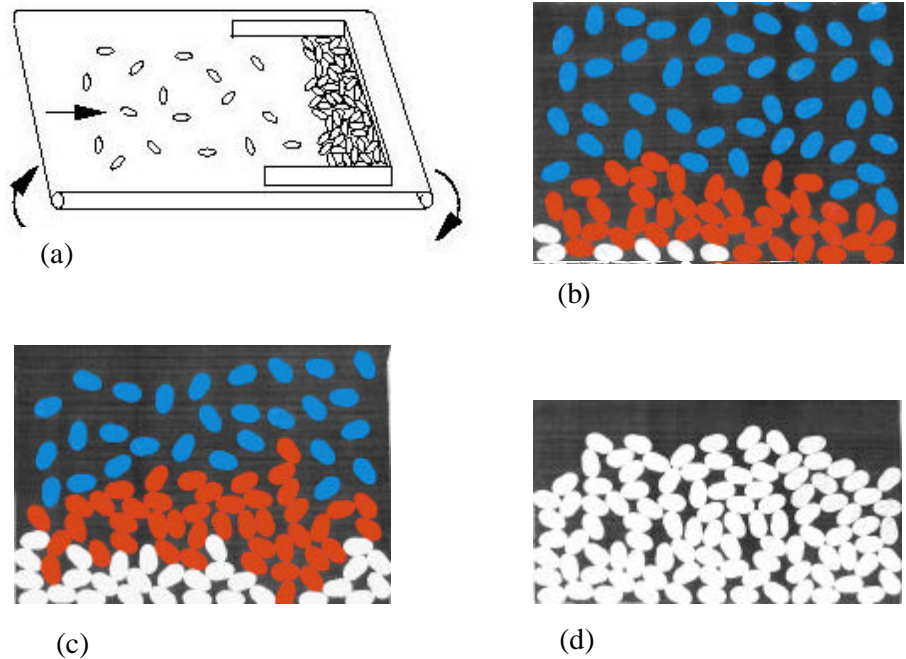

Figure 1. Granular experiments in two dimensions. (a) The basic experimental method is to push a collector towards grains lying on a horizontal surface. (b), (c), (d) show successive stages in the build-up of a pile from grains at one of the higher initial densities studied. Grains which are still in their initial positions have been coloured blue, those which have reached their final position in the pile are shown white, and those which are still reorganising are shown in red. The large red zone reflects the highly cooperative nature of the pile building which is a feature of high initial densities.

Density proved an interesting variable, as shown in figure 2(a). As we raised the density $r_i$ of the initial array of stationary (and non-contacting) grains, the density $r_f$ of the resulting piles decreased. Watching the experiment, the interpretation is that the higher the initial density, the more cooperative is the evolution of the jammed state. The immediate pay-off from the density measurements was the identification of a natural endpoint of our sequence of experiments, where the initial and final densities extrapolate to equality at about $r_c = 0.47$ (area fraction), for our particular shape of grains.

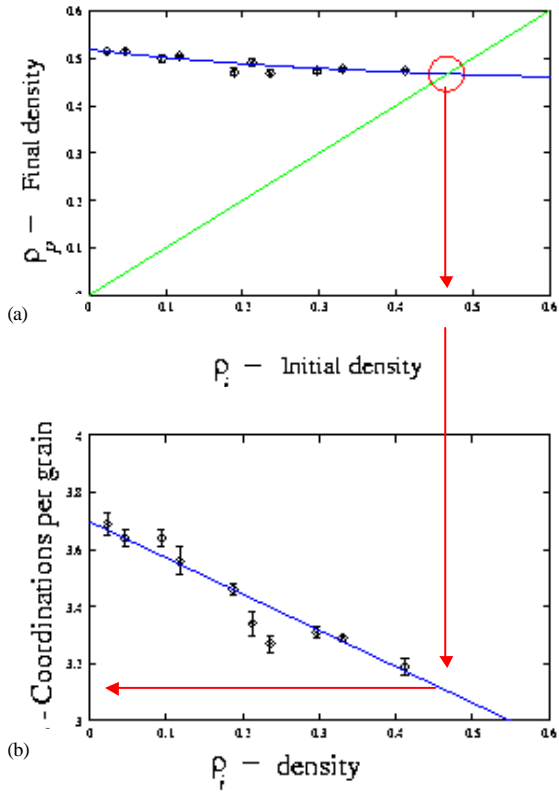

Figure 2. Density and coordination number. (a) The final density (plotted as an area fraction) of our piles showed a weakly inverse trend with the initial density of grains used. This enables us to readily identify a natural end point to the experimental series where the two densities extrapolate to equality. (b) The coordination number decreases with increasing initial density, and the value extrapolated to the end point of the series is about 3.1.

Figure 2 (b) shows the corresponding measurements of coordination number, which we measure as low as 3.2 and which extrapolate to $3.1 \pm 0.1$ at the end point density $r_c$. Thus the Marginally Rigid State appears to be achieved in the limit where initial and final density are matched and the pile formation is at its most cooperative. The rise of coordination number at low initial densities is expected, as the limit of $r_i \to 0$ corresponds to strictly sequential arrival of grains.

Stress Transmission Calculations

Balance of force and torque in continuum mechanics impose only that $\nabla \cdot \underline{\underline{s}} = 0$ and that the stress tensor $\underline{\underline{s}}$ be symmetric, so what else is it that balance of force and torque tells us about transmission of stress in the marginally rigid state? Simple counting shows that we require one further antisymmetric tensor worth of equations to fully determine the stress tensor variation, and as the balance equations are linear in force and torque at the granular level we should expect the 'missing equations' to come out linear in the stress tensor.

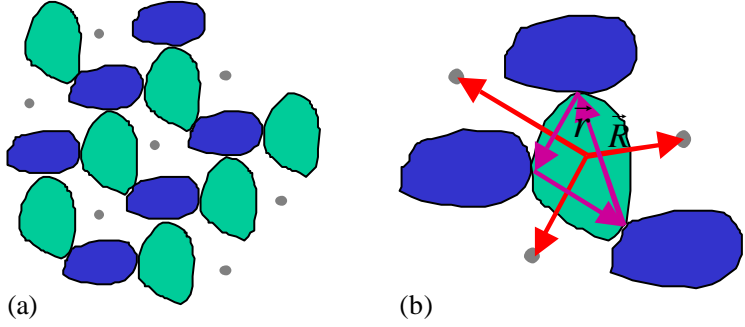

(a)   (b)

Figure 3. (a) The simplest periodic two dimensional lattice of rough grains (supporting static friction) which satisfies the condition of marginal rigidity. (b) Details of the local contact geometry which enter the calculation of coefficients constraining how stress is transmitted through the array.

A variety of granular calculations [4,5,6,7] lead to the simplest possible form, namely local linear relations of the form

$$p_{ijkl}s_{kl} = 0 \quad (1).$$

Fig 3(a) shows the simplest periodic lattice of rough grains satisfying marginal rigidity, and for this (and its three dimensional analogue which has the connectivity of a diamond lattice) the new equations come from requiring that we balance torque on each grain in the unit cell rather than just on the unit cell as a whole.

The fourth rank tensor $p_{ijkl}$ is antisymmetric with respect to interchange of its left two indices and symmetric with respect to interchange of its right indices and depends on the local contact geometry. In the example of figure 3 one can find quite explicit results, with

$$p_{ijkl} = e_{ij}e_{kp} \sum_{\text{roundgrain}} (r_p R_q + R_p r_q) e_{ql} \quad (2)$$

where the vectors $\mathbf{r}$ and $\mathbf{R}$ are indicated in figure 3(b). One has a choice of which grain in the unit cell to use, and they give equal results for $p_{ijkl}$ but of opposite sign.

We have made some progress generalizing the result for $p_{ijkl}$ off-lattice in two dimensions [5], but the dilemma of sign becomes more serious. There is strong anti-correlation between the values for neighbouring grains, and one can show that the local spatial average of the $p_{ijkl}$ goes to zero as the size of region averaged over increases. For the periodic lattice one can consistently adopt a choice of 'even' grains, but this becomes frustrated when odd-numbered coordination rings are allowed. This problem does not appear to arise for smooth grain arrays, and its full implication is not resolved.

Various authors have previously discussed constraints on the stress tensor equivalent to the form $p_{ijkl}s_{kl} = 0$. Cates et al termed this general case OSL [8], and the term 'null stress' has also been applied. An important sub-class is that of Fixed Principle Axes [3,8], which corresponds to $p_{ijkk} = 0$, that is the coupling coefficients being strictly traceless with respect to their right indices: Cates et al showed that the anomalous granular load distribution under a conical pile could be fit well by such a phenomenology [3]. An elegant discussion of linear constraints under less local assumptions has been given by Tkachenko & Witten [9].

Yield and Dissipation

If one admits a finite coefficient of static friction between rough grains, then for periodic arrays it is easy to identify prospective slip planes and the stress threshold at which these are mobilised. This is conventionally encoded in terms of a Yield Function such that for $Y(\underline{\underline{s}}) < 0$ the system is unyielding and (almost) everywhere on the locus $Y(\underline{\underline{s}}) = 0$ slip becomes just possible with a unique orientation specified by the (symmetrised) shear rate tensor being proportional to $\underline{\underline{g}}(\underline{\underline{s}})$.

On the yield surface we can have a creeping yield which we will take the liberty of describing as plastic flow. Adapting the standard plastic flow equations we have the flow field $\underline{u}(\underline{x})$ governed by the following:
$$\partial_i u_j + \partial_j u_i = A(\underline{x},t) g_{ij}(\underline{\underline{s}}) + p_{klij}(\underline{x},t) w_{kl}(\underline{x},t) \qquad (3)$$
where $A(\underline{x},t)$ is a plastic flow amplitude ultimately determined by imposing the yield locus condition $Y(\underline{\underline{s}}) = 0$. The last term on the RHS is new and constitutes a coupling of local grain rolling angular velocity $w_{kl}(\underline{x},t)$ to the macroscopic (and symmetrised) deformation rate; the particular choice of coupling coefficients will be justified further below. The remaining conditions that close the system of equations are force balance $\partial_i s_{ij} = 0$ and our stress constraint relations $p_{ijkl} s_{kl} = 0$.

Coupling the local grain rolling into the macroscopic deformation is a radical step, but introducing some such extra term is necessary if we are to be able to impose the stress constraint relations $p_{ijkl} s_{kl} = 0$ on the yield surface. Not to do so would lead to potentially ill-posed equations, as we could have a system having yielded then drop below the yield surface with stress distribution incompatible with the static equations. Galilean invariance requires that uniform rolling cannot couple to the macroscopic deformation, but this is already assured by the spatial average of $p_{ijkl}$ being zero.

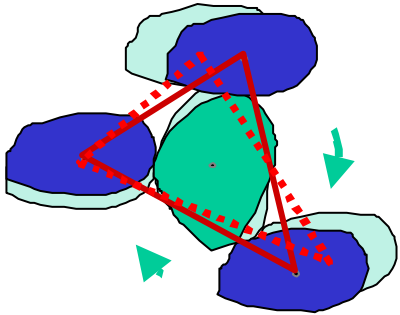

Fig 4. When the central grain rotates maintaining rolling contact with its (non-rotating) neighbours, a non-trivial deformation of the local environment is induced.

The coupling of grain rolling to macroscopic deformation is illustrated in figure 4. Explicit calculation of the mean deformation rate in the triangle of neighbour centres gives

$$\partial_k u_l + \partial_l u_k = \frac{1}{A_\Delta}\int_\Delta (u_k dS_l + u_l dS_k) = \mathbf{w}_{ij} p_{ijkl} \qquad (4)$$

where $p_{ijkl}$ is as given in equation (1).

Consideration of dissipation in creeping flow also leads to the conclusion that the same coefficients $p_{ijkl}(\underline{x},t)$ are common to the stress constraint equations and to the coupling of grain rolling in creeping flow. From the deformation rate of equation (4), the rate of energy dissipation $\dot{\Delta}$ is given by

$$2\dot{\Delta} = A(\underline{x},t) g_{kl}(\underline{\mathbf{S}}) \mathbf{s}_{kl} + \mathbf{w}_{ij}(\underline{x},t) p_{ijkl}(\underline{x},t) \mathbf{s}_{kl} \qquad (5)$$

from which it is apparent that $p_{ijkl}(\underline{x},t)\mathbf{s}_{kl}$ can be interpreted as the force canonically conjugate to grain rolling. The n if one insists that this motion should not be directly dissipative, our stress constraint $p_{ijkl}(\underline{x},t)\mathbf{s}_{kl} = 0$ has to follow. . Similar conclusions are arrived at by Tkachenko and Witten [9].

Renormalisation of Stress Constraints

The property of zero spatial average in the coefficients $p_{ijkl}(\underline{x},t)$ obstructs us obtaining more macroscopic equations by simple pre-averaging. To explore the consequences of this we have in collaboration with R Farr [10] investigated a simple 2x1 block renor malisation scheme as illustrated in figure 5.

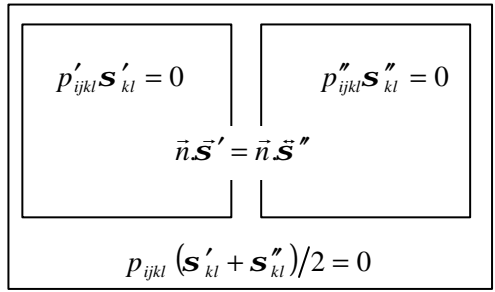

Figure 5. In our renormalisation scheme for the stress constraint coefficients two equal sized blocks (each with uniform $p_{ijkl}$ and $\underline{\mathbf{s}}$) are joined with matching of boundary forces, and we seek the corresponding constraint obeyed by the spatial average stress.

In two dimensions, antisymmetry renders the left two indices of $p_{ijkl}$ redundant and the renormalisation simply yields $p_{kl} = B\left(p'^{-1}_{tt} p'_{kl} + p''^{-1}_{tt} p''_{kl}\right)$ where $t$ denotes the direction parallel to the joining surface and $B$ is a free parameter. This extends to higher dimensions but the inversion indicated is less trivial.

When this renormalisation is iterated we find three classes of Fixed Point as follows.

1. *Constant Fixed Points:* $p_{ijkl}$ = any constant tensor

The significance of these is unclear, as we start with zero average $p_{ijkl}$ at grain level, but we should not dismiss them as the RG transformation not linear.

2. *Pressureless fixed points*

Here (one of) the $p_{..kl..}$ constrains the stress to be traceless, which is clearly a transitive property under the stress averaging. These fixed points are linearly stable to fluctuations in $d=2$, by both analytic calculations and numerical investigation by R Farr [10]. Unfortunately pressureless stress is incompatible with cohesionless granular matter, as at least one principal stress must be positive.

3. *Fixed Principle Axes fixed points*

These correspond to all the coefficients being traceless with respect to their right hand indices, i.e. $p_{ijkk} = 0$ for all $i,j$ which is clearly transitive under our renormalisation of the $p_{ijkl}$. We find such fixed points can be neutrally stable with respect to the perturbation of small isotropic parts in $p_{ijkl}$, by analytic calculations in $d=2$ and with numerical indications [10] in $d=3$.

The stability of the unphysical pressureless fixed points and the only marginal stability of the FPA fixed points are discouraging. However it should be noted that we are not assured to have exhausted the fixed points, and of course that the renormalisation itself is a crude model.

Adjustment to the Stress Constraints

Missing in all the above discussions is how a sample will accommodate a change in external loading: to what extent will this entail deformation and adjustment of the $p_{ijkl}$, as opposed to simply propagating the new forces (adjusting $\underline{\underline{s}}$) through the pre-existing geometry. Evidently a mixture of both is generally involved, but in this paper we would like to suggest a qualitative distinction might arise between rough and smooth grains.

Smooth grains

Rough grains

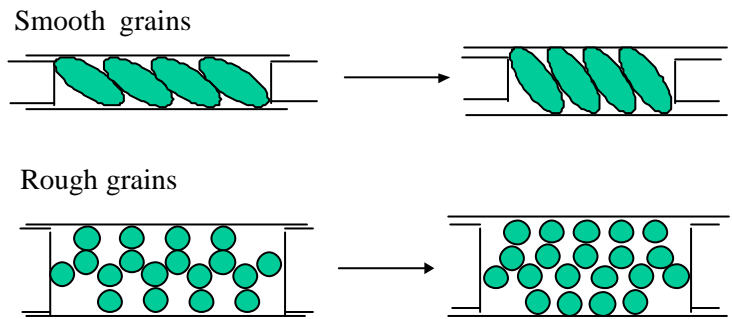

Figure 6. The above are minimal arrays of respectively rough and smooth grains conforming to the conditions of marginal rigidity, and the manner in which they accommodate a uniaxial compression is shown. The smooth grains do so in a manner

which simply extends to larger arrays. By contrast the rough grains exhibit excess rearrangement near the walls, because the amount of grain rolling required to accommodate compression of the central spine overshoots the amount of roll to commensurately compress the top and bottom of the array.

Figure 6 illustrates the response of minimal marginally rigid arrays in compression. For smooth grains it is clear that deformation can penetrate uniformly into the bulk of a periodic array, whilst for rough grains the response is inhomogeneous and biassed towards the walls. In simple shear (which a periodic array of rough grains facilitates by rolling) it can be shown that the relative shear displacement across the wall layer is of the same order as the total imposed displacement.

These observations prompt us to conjecture that for smooth grain assemblies external perturbations can lead to deformations which penetrate the bulk, whereas for rough grain assemblies they lead to preferential rearrangement at the surface which may lead to new contacts and effectively block deformation penetrating the bulk. The two cases might be distinguished as marginally fluid and marginally rigid respectively. Yield would then amount to a forced transition from the second case to the first, though local sliding. We should emphasise that this paragraph remains purely conjectural.

Acknowledgements:
This work was inspired by the ideas of SF Edwards on Granular materials, and we also benefited through input from D Grinev, R Farr, C Thornton and J Melrose. R Blumenfeld was funded by the EPSRC through research grant GR/L5975.